\documentclass[useAMS,usenatbib]{mn2e}
\usepackage{psfig}
\usepackage{epsfig}
\usepackage{longtable}
\usepackage{times}
\usepackage{lscape}

\newbox\grsign \setbox\grsign=\hbox{$>$} \newdimen\grdimen
\grdimen=\ht\grsign
\newbox\simlessbox \newbox\simgreatbox \newbox\simpropbox
\setbox\simgreatbox=\hbox{\raise.5ex\hbox{$>$}\llap
     {\lower.5ex\hbox{$\sim$}}}\ht1=\grdimen\dp1=0pt
\setbox\simlessbox=\hbox{\raise.5ex\hbox{$<$}\llap
     {\lower.5ex\hbox{$\sim$}}}\ht2=\grdimen\dp2=0pt
\setbox\simpropbox=\hbox{\raise.5ex\hbox{$\propto$}\llap
     {\lower.5ex\hbox{$\sim$}}}\ht2=\grdimen\dp2=0pt

\def\Rsun{\hbox{$\rm\thinspace R_{\odot}$}}
\def\Msun{\hbox{$\rm\thinspace M_{\odot}$}}
\def\Mjup{\hbox{$\rm\thinspace M_{\rm Jup}$}}
\def\Rjup{\hbox{$\rm\thinspace R_{\rm Jup}$}}

\def\mjup{\mbox{$M_\textrm{\tiny Jup}$}}
\def\msun{\mbox{$M_\odot$}}

\def\ga{\mathrel{\hbox{\rlap{\hbox{\lower4pt\hbox{$\sim$}}}{\raise2pt\hbox{$>$}}}}}\def\la{\mathrel{\hbox{\rlap{\hbox{\lower4pt\hbox{$\sim$}}}{\raise2pt\hbox{$<$}}}}}

\begin{document}

\title[NLTT\,5306]{NLTT\,5306: The shortest Period Detached White Dwarf $+$ Brown Dwarf Binary}

\author[P.\,R. Steele et~al. ]
{
P.\,R. Steele$^1$, R.\,P. Saglia$^1$, M.\,R. Burleigh$^2$, T.\,R. Marsh$^3$, B.\,T. G\"ansicke$^3$, K. Lawrie$^2$,\newauthor M. Cappetta$^1$, J. Girven$^3$ and R. Napiwotzki$^4$ \\
$^1$ Max-Planck-Institut f\"{u}r extraterrestrische Physik, Giessenbachstrasse, 85748, Garching, Germany\\
$^2$ Department of Physics and Astronomy, University of Leicester, University Rd., Leicester, LE1 7RH, UK\\
$^3$ Department of Physics, University of Warwick, Coventry, CV4 7AL, UK\\
$^4$ Centre for Astrophysics Research, University of Hertfordshire, College Lane, Hatfield, AL10 9AB, UK\\ 
}

\date{Submitted 2012 October 26, Accepted 2012 December 12}

\maketitle

\begin{abstract}
We have spectroscopically confirmed a brown dwarf mass companion to the hydrogen atmosphere white dwarf NLTT\,5306. The white dwarf's atmospheric parameters were measured using Sloan Digital Sky Survey and X-Shooter spectroscopy as $T_{\rm eff}=7756\pm35$K and log(g)$=7.68\pm0.08$, giving a mass for the primary of $M_{\rm WD}=0.44\pm0.04$\,\msun\, at a distance of $71\pm4$\,pc with a cooling age of 710$\pm$50\,Myr. The existence of the brown dwarf secondary was confirmed through the near-infrared arm of the X-Shooter data and a spectral type of dL4-dL7 was estimated using standard spectral indices. Combined radial velocity measurements from the Sloan Digital Sky Survey, X-Shooter and the Hobby-Eberly Telescope's High Resolution Spectrograph of the white dwarf gives a minimum mass of $56\pm3$\,\mjup\ for the secondary, confirming the substellar nature. The period of the binary was measured as $101.88\pm0.02$\,mins using both the radial velocity data and $i'$-band variability detected with the INT. This variability indicates 'day' side heating of the brown dwarf companion. We also observe H$\alpha$ emission in our higher resolution data in phase with the white dwarf radial velocity, indicating this system is in a low level of accretion, most likely via a stellar wind. This system represents the shortest period white dwarf $+$ brown dwarf binary and the secondary has survived a stage of common envelope evolution, much like its longer period counterpart, WD\,0137$-$349. Both systems likely represent bona-fide progenitors of cataclysmic variables with a low mass white dwarf and a brown dwarf donor.   

\end{abstract}


\section{Introduction}

As the descendants of high mass ratio binaries, brown dwarf (BD) companions to white dwarfs (WDs) enable investigation of one extreme of binary formation and evolution, including the known deficit of BD companions to main sequence stars \citep{mccarthy04,grether06}. The BDs can be directly detected relatively easily at all separations, since they dominate the spectral energy distribution at near- to mid-infrared (IR) wavelengths, in stark contrast to main sequence $+$ substellar pairs. 

As detached companions to WDs, BDs are rare ($\sim0.5$\%,\,\citealt{steele11,girven11}). Only a handful of such systems have thus far been spectroscopically confirmed e.g. GD165 (DA$+$dL4, \citealt{becklin88}), GD\,1400 (DA$+$dL6-7, \citealt{farihi04,dobbie05}), WD\,0137$-$349 (DA$+$dL8, \citealt{maxted06,burleigh06}), PHL\,5038 (DA$+$dL8; \citealt{steele09}), and LSPM\,1459$+$0857 (DA$+$T4.5, \citealt{dayjones11}). GD\,165, PHL\,5038 and LSPM\,1459$+$0857 can be classed as widely orbiting with projected separations of 120\,AU, 55\,AU and 16500-26500\,AU respectively. WD\,0137$-$349 and GD\,1400 have much shorter orbital periods of 116\,mins and $\sim10$\,hrs \citep{burleigh11} respectively, and have both undergone a common envelope evolution. 

These two distinct populations are thought to be the outcome of stellar evolution; the wide pairs where the secondary has migrated outwards due to the mass loss of the WDs progenitor \citep{farihi06,nordhaus10}, and the close systems in which the secondary has survived a stage of common envelope evolution and may eventually lead to the formation of a cataclysmic variable (CV) \citep{politano04}. In these close binaries, the BD is expected to be irradiated by the WD's high UV flux, leading to substantial differences in the 'day' and 'night' side hemispheres. These systems can additionally be used for testing models of irradiated 'hot Jupiter' atmospheres (e.g. HD\,189733b, \citealt{knutson07}).

NLTT\,5306 ($=$SDSS\,J\,013532.98$+$144555.8) was first identified as a candidate WD$+$BD binary in \cite{steele11} and \cite{girven11}. The former used an estimate of the WD's atmospheric parameters ($T_{\rm eff}=8083\pm22$ and log$g=8.08\pm0.04$; \citealt{eisenstein06}) in combination with cooling models for hydrogen atmosphere (DA) WDs \citep{holberg06,kowalski06,tremblay11,bergeron11} to predict the star's near-infrared photometry. A comparison was then made with the UKIDSS observations identifying NLTT\,5306 as having a near-infrared excess consistent with a red companion. The SDSS and UKIDSS magnitudes are given in Table~\ref{magnitudes}. Further fitting of the photometry yielded an estimated spectral type of dL5 for the secondary, with a mass of $58\pm2M_{\rm Jup}$ at a distance of $60\pm10$\,pc. It should be noted that this spectroscopic mass estimate is model dependent, calculated by interpolating the Lyon group atmospheric models \citep{chabrier00,baraffe02} given an estimated age for the WD and temperature for the BD. The system was unresolved with an upper limit on the projected separation of $<57$\,AU between components.

The structure of this paper is as follows; In Section~2 we describe the observations and their reduction. In Section~3 we describe the analysis of the data, starting with the optical light curve, followed by the optical and NIR spectroscopy of the WD and BD, and finally the radial velocity. In Section~4 we discuss the implications of the results and state our conclusions. 

\section{Observations and Data Reduction}

\begin{table}
\caption{SDSS and UKIDSS magnitudes of NLTT\,5306.} 
\label{magnitudes} 
\centering   
\begin{tabular}{c c c c c}    
\hline\hline    
Band & Magnitude \\   
\hline   
$u'$ & $17.51\pm0.01$ \\
$g'$ & $17.03\pm0.00$ \\
$r'$ & $16.95\pm0.00$ \\
$i'$ & $16.96\pm0.01$ \\
$z'$ & $17.00\pm0.01$ \\
$Y$ & $16.49\pm0.01$ \\
$J$ & $16.24\pm0.01$ \\
$H$ & $15.86\pm0.01$ \\
$K$ & $15.56\pm0.02$ \\
\hline                             
\end{tabular}
\end{table}

\subsection{INT Optical Photometry}

NLTT\,5306 was observed photometrically for two hours (80$\times$90s exposures) in the Sloan i'-band on the night of 2009 October 23 with the Wide Field Camera (WFC) on the 2.5\, Isaac Newton Telescope (INT) in La Palma, Spain. The data were reduced using the INT Wide Field Survey (WFS) pipeline \citep{irwin01} developed by the Cambridge Astronomical Survey Unit. For a detailed description of the reduction process, see \cite{irwin07}. In brief a standard CCD reduction was performed by correcting for the bias, trimming the frames, correcting non-linearity, flat-fielding and correcting for the gain. The flux was measured in each observation using aperture photometry and the result converted to magnitudes using nightly zero-point estimates based on standard star field observations \citep{irwin07}.

\subsection{X-Shooter Spectroscopy}

\begin{table*}
\caption{Measured atmospheric parameters for NLTT\,5306\,A from various spectroscopic observations.} 
\centering   
\begin{tabular}{c c c c c c c}
\hline                     
Telescope/Instrument & $T_{\rm eff}$ (K) & log$g$ & $M_{/rm WD}$ (\Msun) & d (pc)  \\     
\hline
SDSS 1 & $7641\pm48$ & $7.61\pm0.10$ & $0.39\pm0.05$ & $72\pm4$  \\ 
SDSS 2 & $7729\pm7$  & $7.67\pm0.05$ & $0.42\pm0.02$ & $68\pm2$  \\ 
SDSS 3 & $7729\pm49$ & $7.71\pm0.10$ & $0.44\pm0.05$ & $68\pm4$  \\
VLT$+$X-Shooter & $7925\pm15$ & $7.74\pm0.02$ & $0.47\pm0.01$ & $70\pm1$  \\
\hline                             
\end{tabular}
\label{param2} 
\end{table*}

NLTT\,5306 was observed using X-Shooter \citep{dodorico06} mounted at the VLT-UT2 telescope on the night of 2010 September 5. X-Shooter is a medium-resolution spectrograph capable of observing using 3 independent arms simultaneously; the ultraviolet (UVB), optical (VIS) and the near-infrared (NIR) arms covering a wavelength range of 0.3-2.5$\mu$m. For our observations we used slit widths of 0.8$\arcsec$, 0.9$\arcsec$ and 0.9$\arcsec$ in the UVB, VIS and NIR arms respectively. Exposure times for each arm were 1200 s in the UVB, 1200 s in the VIS and 12 x 150 s in the NIR. We nodded between each exposure along the NIR slit to improve sky subtraction. This gave us a total of 4 exposures in each arm.

Reduction of the raw frames was carried out using the standard pipeline release of the ESO X-Shooter Common Pipeline Library recipes (version 1.3.7) within \textsc{GASGANO}\footnote{http://www.eso.org/sci/software/gasgano}, version 2.4.0. The standard recipes were used with the default settings to reduce and wavelength calibrate the 2-dimensional spectrum for each arm. The extraction of the science and spectrophotometric standard and telluric spectra were carried out using \textsc{APALL} within \textsc{IRAF}. The instrumental response was determined by dividing the associated standard star by its corresponding flux table. We also used this method to apply the telluric correction.

Finally, the spectra were flux calibrated using the SDSS and UKIDSS magnitudes (Table~\ref{magnitudes}).

\subsection{SDSS Spectroscopy}

NLTT\,5306 was observed on multiple occasions by the SDSS (=SDSS~J013532.97+144555.9). From the SDSS archive we extracted a total of 17 spectra taken over the period from 2000 December 1 to 20. All but three of these spectra had exposure times of 15 minutes; the three spectra acquired on December 4 were exposed for 20 minutes. 

\subsection{High Resolution Spectroscopy with the Hobby-Eberly Telescope} 

NLTT\,5306 was observed using the High Resolution Spectrograph (HRS; Tull 1998) on the Hobby-Eberly Telescope (HET; Ramsey et al. 1998) on the nights of 2010 December 6 \& 10, 2011 January 16 \& 23, and 2011 February 8. The ephemeris of the system was unknown at the time and so 6 random observations were taken in order to establish if the primary has a measurable radial velocity variation, and to then estimate an orbital period. Each observation was split into 2 separate exposures of 1320 s. A ThAr lamp exposure was taken both before and after the science observations in order to aid wavelength calibration.

The cross disperser setting was '316g5936' corresponding to a wavelength range of 4076-7838\AA, in order to cover the H$\alpha$, H$\beta$ and $H\gamma$ Balmer lines. This gives a spectral resolving power of $R=\lambda/\delta\lambda=15,000$. Two sky fibers were used to simultaneously record the sky background. 

Reduction of the raw frames was carried out using standard routines in \textsc{IRAF}. In brief, bias and flat frames were combined and used to correct the science frames. The extraction of the science spectra were carries out using \textsc{APALL} within \textsc{IRAF}. The sky spectrum was extracted in the same way as the science, simply by shifting all the apertures by a set amount to cover the parallel sky fiber. The extracted sky spectrum was then scaled so that the sky lines matched the strength of the corresponding lines in the science spectra. Particular attention was payed to the order containing the H$\alpha$ absorption. This was then subtracted from the science spectra. Finally the sky subtracted science spectra were normalised using the \textsc{CONTINUUM} package within \textsc{IRAF}.

\section{Analysis}

\subsection{Optical Light Curve}
\label{int}

\begin{figure}
\centering
\includegraphics[width=9.0cm]{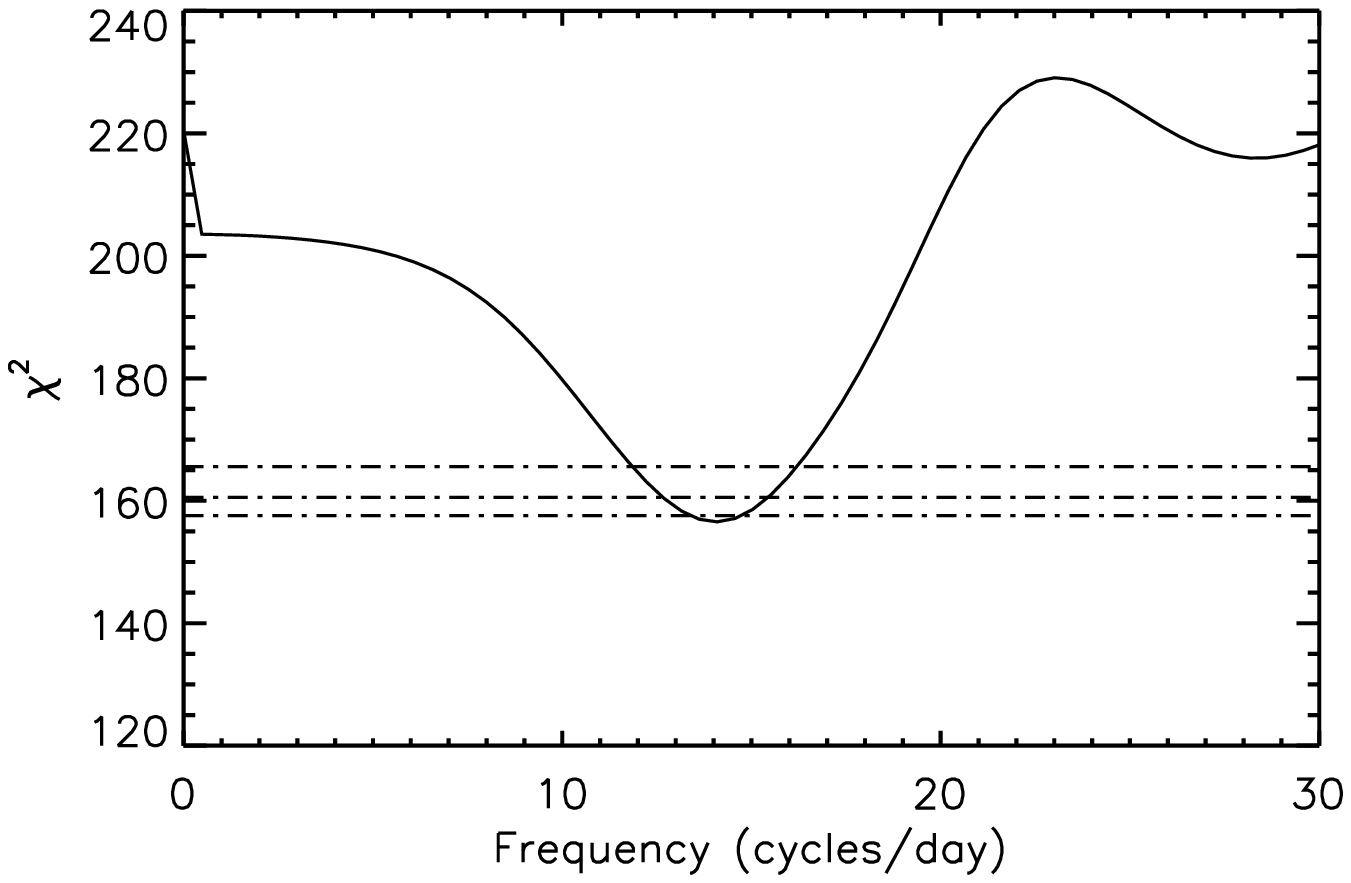} 
\caption{Floating-mean periodogram for the INT $i$'-band photometry of NLTT\,5306. The global minimum is located at a frequency of 14.1$^{+\,0.9}_{-\,1.4}$\,cycles/day.}
\label{int_pd}
\includegraphics[width=6.0cm,angle=270]{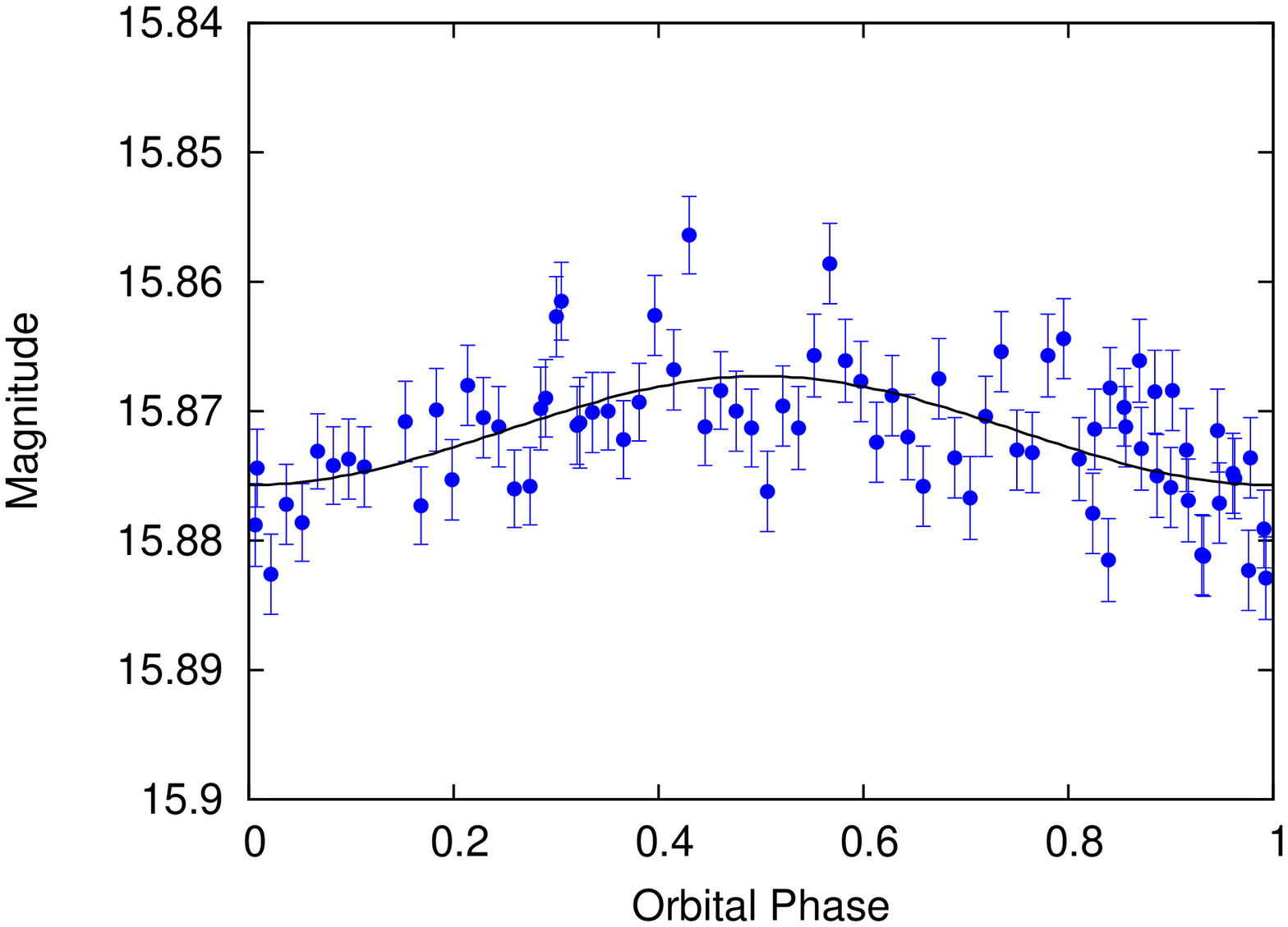} 
\caption{Phase folded INT i'-band light curve of NLTT\,5306 showing a peak-to-peak variability of $\sim1$\% with a period of 102.2$^{+\,11.3}_{-\,6.4}$\,min.}
\label{int_lc}
\end{figure}

The INT $i$'-band light curve of NLTT\,5306 (Figure~\ref{int_lc}) shows low-level photometric variability with a peak-to-peak amplitude of $\approx0.8$\%. The analysis leading to this result is outlined below.    

A `floating-mean' periodogram was used to search for periodicity in the target (\citealt*{cumming99}). This method involves fitting the time-series data with a sinusoid plus a constant $A$ in the form of: $$A+B \textrm{sin}[2\pi f(t-t_0)]\,\,\,(1)$$ where $f$ is the frequency and $t$ is the time of observation. The resulting periodogram is a $\chi^2$ plot of the fit with frequency (Figure~\ref{int_pd}). The errors associated with the best-fitting frequency were estimated as the 2$\sigma$ frequency range from the global minimum (corresponding to a change in $\chi^2$ of four, assuming only one useful fitted parameter).

To evaluate the significance of the best-fitting period, a false alarm probability (FAP) was estimated using 100000 Monte Carlo trials and an analytical approach for comparison. Fake light curve datasets were generated for the Monte Carlo tests at the same timings as the observations with the mean magnitude as the observed data. Random Gaussian noise was then added to the flux distributed with the same variance as the observed magnitudes. The FAP was determined from the number of trials where the maximum power in the periodogram (from the fake dataset) exceeded the maximum power from the observed dataset. A significant detection threshold was set at 1\% (a of FAP\,$\leq0.01$). The analytical probability was determined using the equation given in Table 1 in \cite{zechmeister09} for the residual variance normalisation (also see Appendix B in \citealt{cumming99}). Further details on the significance tests can be found in \cite{cumming99} and Lawrie et al. (2012, in prep.).

A global minimum in the periodogram is found at a frequency of 14.1$^{+\,0.9}_{-\,1.4}$\,cycles/day, a period of 102.2$^{+\,11.3}_{-\,6.4}$\,min. A fitted sine wave to the data gives a reduced $\chi^2$ of 2.07 ($\chi^2$ of 157 over 76 degrees of freedom (dof)), while a constant fit to the data gives a reduced $\chi^2$ of 2.92 ($\chi^2$ of 231 over 79 dof). The FAP statistics are well within the limit for a significant detection, with a FAP of $<$0.001 from the Monte Carlo tests and a FAP of $5\times10^{-5}$ from the analytical estimation. This suggests that it is significantly unlikely that the variability seen in the light curve is due to noise fluctuations alone.

The probability that this system has an alignment that would result in an eclipse as viewable from Earth is $\sim20$\%, with an eclipse duration of $\sim7$\,mins \citep{faedi11}. No eclipse or grazing transit is immediately obvious in the phase folded light curve of NLTT\,5306 (Figure~\ref{int_lc}). Given the total time between exposures of 90s and the coverage of $\sim1.5$\, orbital periods, it seems unlikely that we would have missed an eclipse. 

\subsection{White Dwarf Spectroscopy}

\begin{figure}
\centering
\includegraphics[width=5.5cm,angle=270]{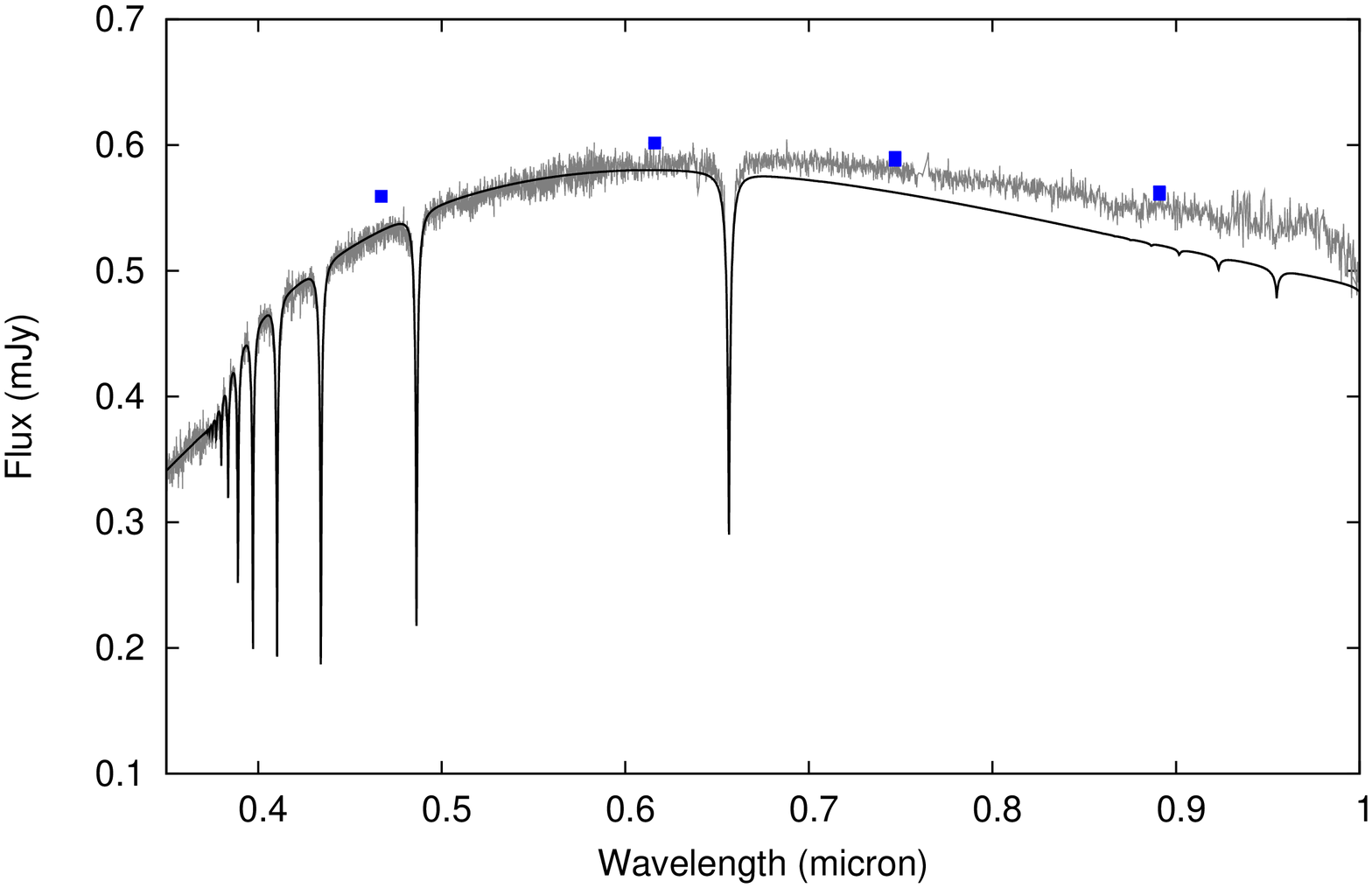}
\caption{X-Shooter spectrum (grey) of WD\,0132$+$142 covering the UVB and VIS arms where the WD primary dominates. SDSS $ugriz$ photometry is overplotted (squares). The WD model SED (black) begins to diverge from the observed spectrum at longer wavelenghts due to the added flux from the secondary.}
\label{xshooter_wd}
\includegraphics[width=5.5cm,angle=270]{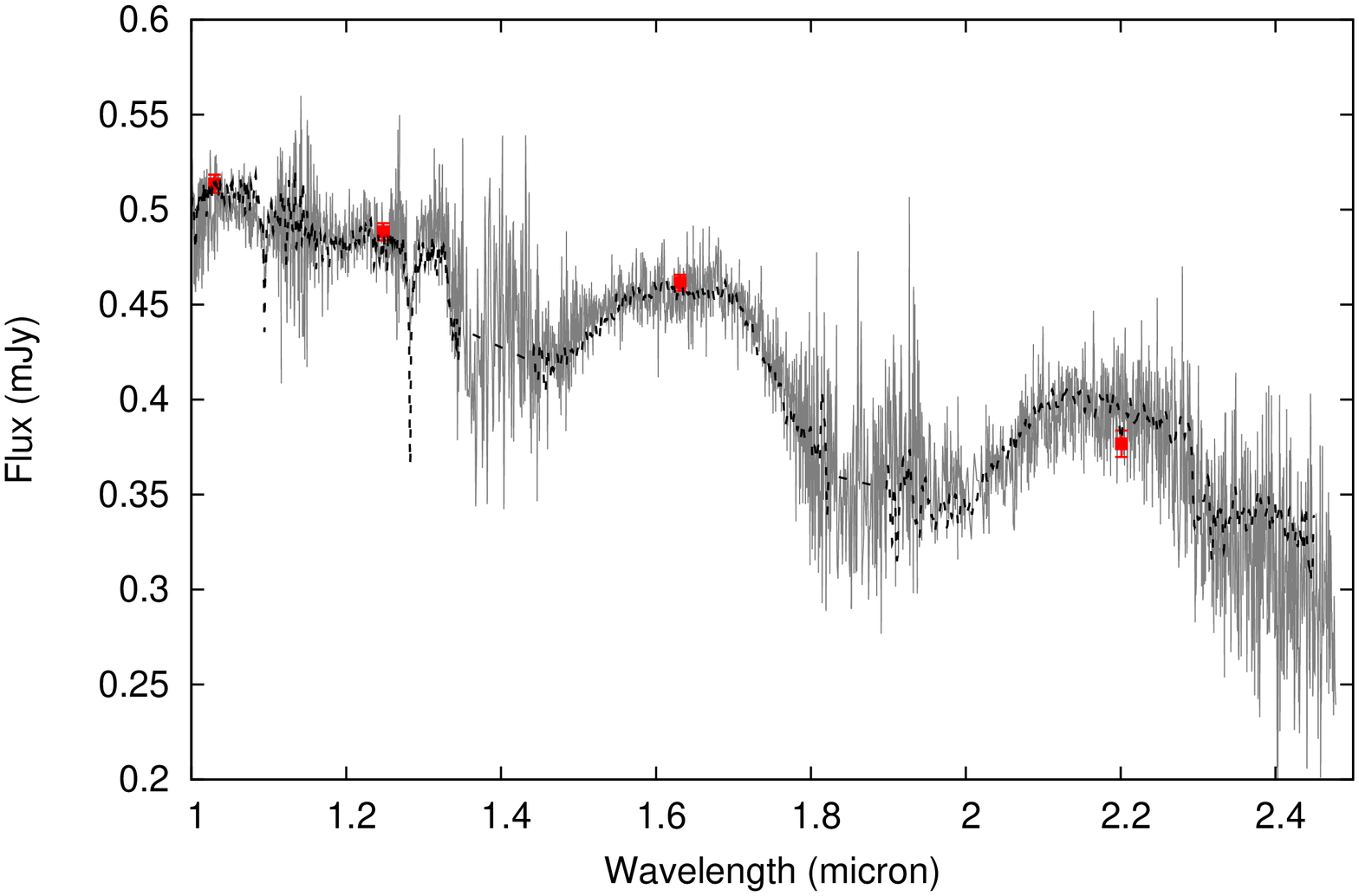}
\caption{X-Shooter Spectrum (grey) of WD\,0132$+$142 covering the NIR arm where the BD secondary dominates. UKIDSS $YJHK$ photometry is overplotted (squares) as well as a composite WD+dL5 model (black dashed) for visual comparison.}
\label{xshooter_bd}
\end{figure}

\begin{figure}
\centering
\includegraphics[width=6.0cm,angle=270]{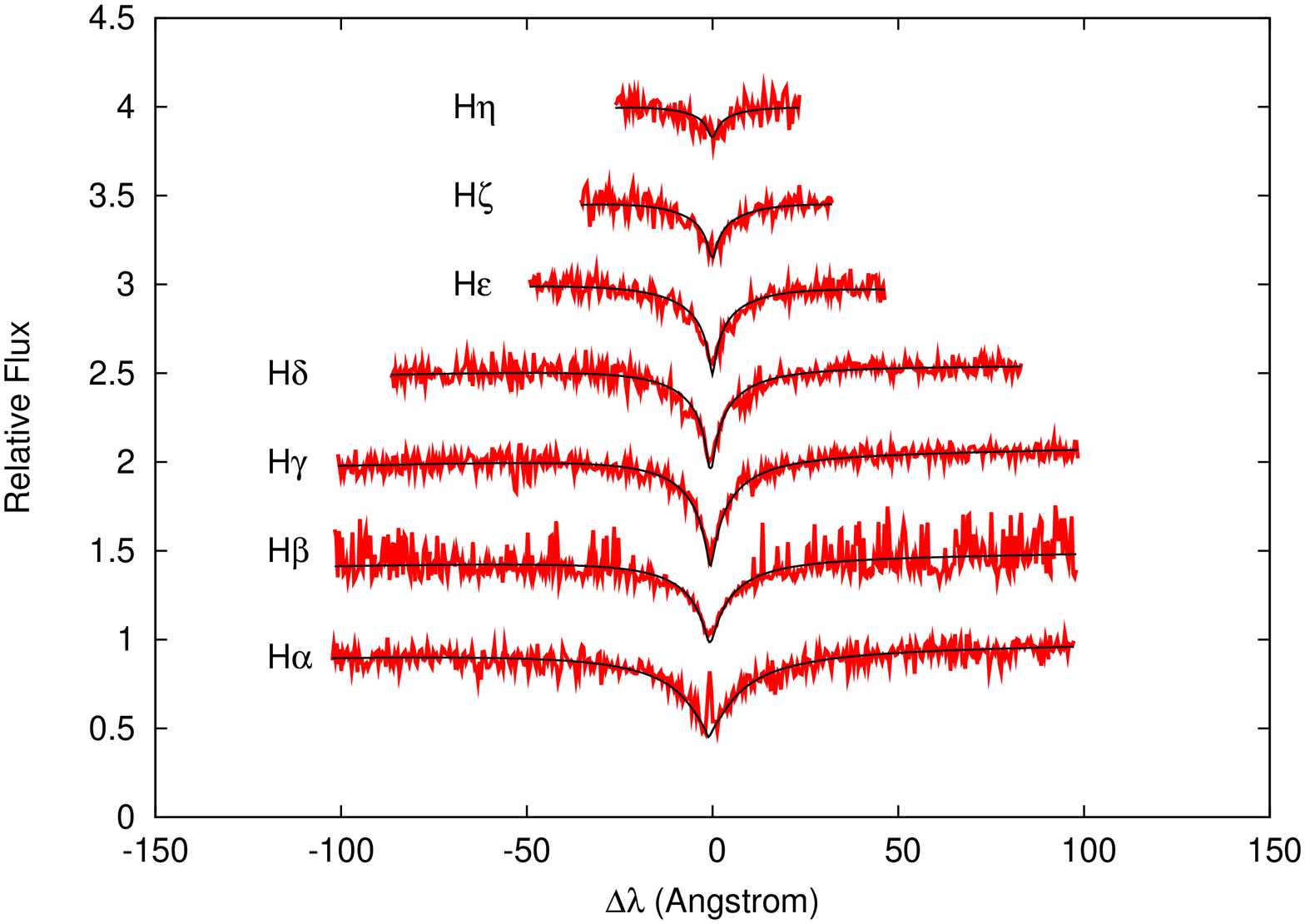}
\caption{Balmer line fit (black) of NLTT\,5306\,A using the X-Shooter spectrum (grey). The emission line seen at the core of H$\alpha$ is likely due to  accretion via a stellar wind. The region of the emission line was excluded from the fit.}
\label{xshooter_fit}
\centering
\includegraphics[width=5.5cm,angle=270]{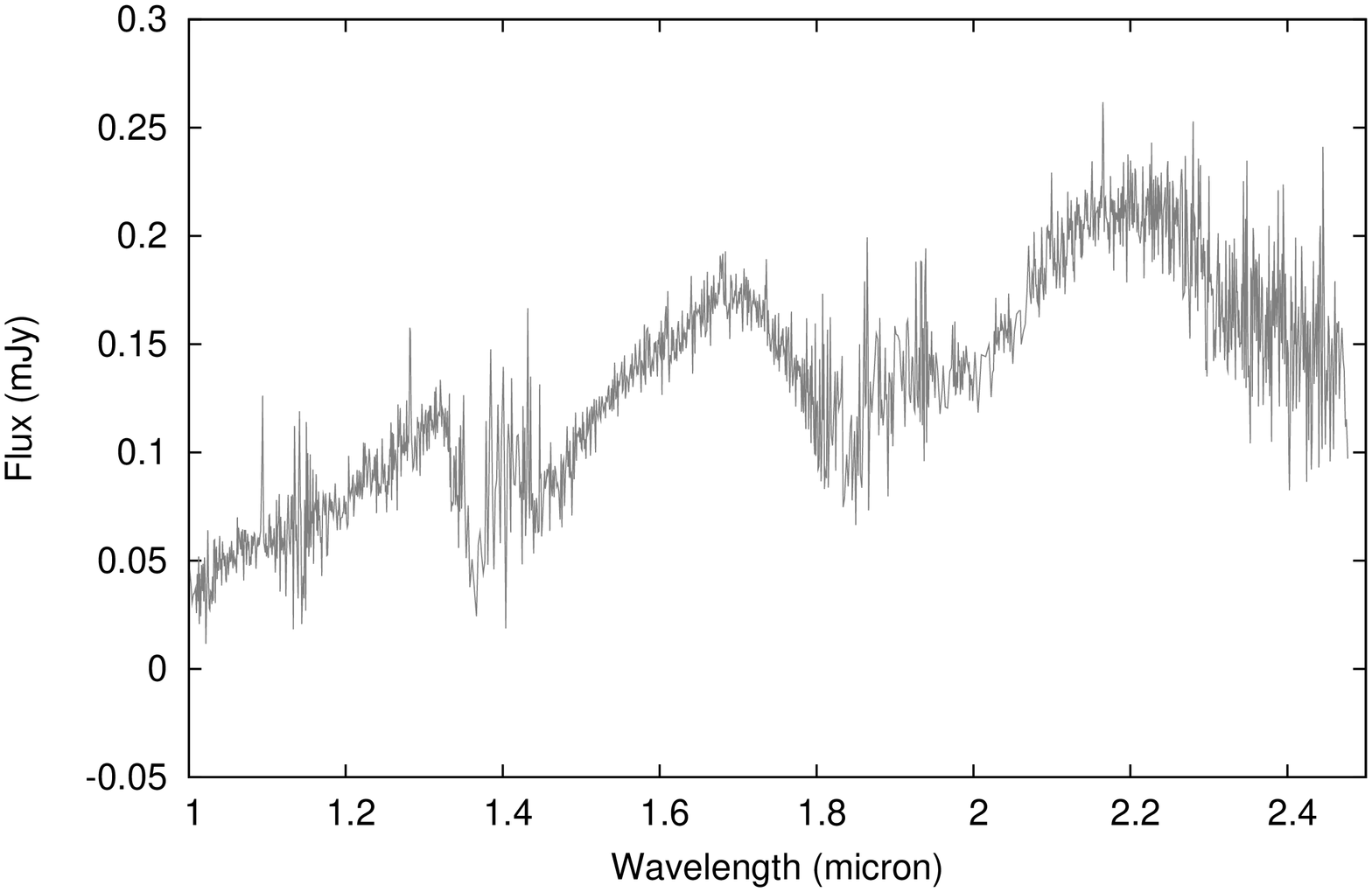}
\caption{X-Shooter spectrum of NLTT\,5306\,B (grey) created by subtracting the WD model spectrum calculated using the measured atmospheric parameters of the primary.}
\label{wd0132b_bd}
\end{figure}

Figures~\ref{xshooter_wd} and ~\ref{xshooter_bd} show the extracted X-Shooter spectrum for NLTT\,5306 separated into the UVB/VIS and NIR arms. This clearly shows the WD primary dominating the optical wavelengths and the BD companion dominating the near-infrared wavelengths. A closer inspection of the H$\alpha$ absorption line in individual exposures revealed the presence of line emission close to the line centre (Figure~\ref{xshooter_fit}). 

Figure~\ref{xshooter_fit} and Table~\ref{param2} shows the results of fitting the Balmer series with atmospheric models of hydrogen atmosphere DA WDs \citep{koester08}, for both the SDSS and X-Shooter spectra. The average of these values yields and effective temperature and surface gravity of $T_{\rm eff}=7756\pm35$K and log$g=7.68\pm0.08$ respectively. We interpolated these values over a grid of synthetic colours and evolutionary sequences of DA WDs\footnote{http://www.astro.umontreal.ca/$\sim$bergeron/CoolingModels/} to calculate a mass of $M_{\rm WD}=0.44\pm0.04$\Msun\ and a distance of $71\pm4$\,pc. The results are summarised in Table~\ref{param_wd}. The region within H$\alpha$ containing the core emission was excluded from this fit. A model spectrum calculated using these values is overplotted in Figure~\ref{xshooter_wd}.

It should be noted that the spectroscipic fit may have lead to an overestimate of the WD's mass, due to the difficulty in modelling the hydrogen line profiles below $T_{\rm eff}\approx12000-13000$\,K. The \citet*{tremblay11} analysis of SDSS WDs suggest an overestimate at the level of 10\%, although this is largely based on WDs with a mass of around 0.6\,\Msun. A lower WD mass would decrease the minimum value for the mass of the secondary, relaxing the need for the system to have a high inclination for the mass to be consistent with the spectral type of dL4-7 and given the non-detection of an eclipse or grazing transit (Section~\ref{int}).


\subsection{Brown Dwarf Spectroscopy}

The model WD spectrum for NLTT\,5306\,A was subtracted from the observed X-Shooter spectrum in order to estimate the spectral energy distribution of solely the secondary, NLTT\,5306\,B (Figure~\ref{wd0132b_bd}). 

In order to estimate a spectral type for the secondary we have calculated the H$_2$O$^{\rm b}$ and H$_2$O$^{\rm c}$ indices, defined by: $${\rm H}_{2}{\rm O}^{\rm b}={\rm F}(1.48)/{\rm F}(1.60)\,\,\,(2)$$ $${\rm H}_{2}{\rm O}^{\rm c}={\rm F}(1.80)/{\rm F}(1.70)\,\,\,(3)$$ where F is the average flux in a band, with $\pm0.01\mu$m, centred on the specified wavelength. \cite{burgasser02} state that the H$_2$O$^{\rm b}$ index in particular shows a linear relation with spectral type given by: $$ {\rm SpT}=(12.6\pm0.9)-(26.7\pm0.6){\rm H}_{2}{\rm O}^{\rm b}\,\,\,(4)$$ where SpT=0 at dT0 and SpT=-4 at dL5. This relation holds for spectral types between dM5 (-14) and dT8 (8).

\begin{figure}
\centering
\includegraphics[width=8.0cm]{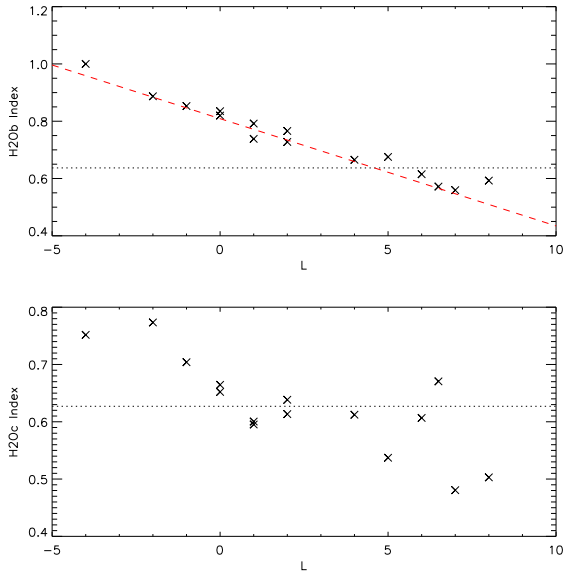}
\caption{H$_2$O$^{\rm b}$ and H$_2$O$^{\rm c}$ Indices as measured for L-dwarfs from the NIRSPEC Infrared Archive (crosses). Ths short-dashed lines shows the same indices measured for NLTT\,5306\,B. These suggests the spectral type in the approximate range dL4-dL7. The H$_2$O$^{\rm b}$ index (Upper panel) is the better indicator of the spectral type. Equation~4 is overplotted in the upper panel (long-dashed) to show the linear relationship between spectral type and the H$_2$O$^{\rm b}$ index.}
\label{xshooter_water}
\end{figure}

Figure~\ref{xshooter_water} shows these indices as measured for L-dwarfs from the NIRSPEC infrared archive\footnote{http://www.astro.ucla.edu/~mclean/BDSSarchive/}, as well as the H$_2$O$^{\rm b}=0.637$ and H$_2$O$^{\rm c}=0.627$ as  measured for NLTT\,5306\,B. Substituting the former value into equation~4 gives SpT=$-4.4\pm1.3$, corresponding to a spectral type between dL4-dL7. This is consistent with the original photometry based spectral type estimate of dL5 in \citet{steele11}. 



\subsection{Radial Velocity}

\begin{table}
\caption{Observation times and measured radial velocities from the Balmer absorption lines of NLTT\,5306. The SDSS data have been shifted to the mean systemic velocity of the X-Shooter and HET observations.} 
\centering   
\begin{tabular}{l c c}
\hline                     
Telescope/Instrument & HJD & RV (kms$^{-1}$)  \\     
\hline\hline
SDSS &  2451879.73722 & -39.06$\pm$14.4 \\
     & 2451879.74935 & -57.86$\pm$14.6 \\
     & 2451879.78591 & 58.64$\pm$17.7 \\
     & 2451882.75429 & 51.24$\pm$9.8 \\ 
     & 2451882.77006 & 29.84$\pm$9.9 \\ 
     & 2451882.78609 & -46.46$\pm$11.5 \\
     & 2451884.67157 & 48.14$\pm$11.0 \\
     & 2451884.68378 & -3.96$\pm$11.7 \\
     & 2451884.69593 & -52.26$\pm$10.9 \\
     & 2451884.70808 & -17.46$\pm$12.2 \\
     & 2451884.72014 & 10.94$\pm$12.2 \\
     & 2451884.73226 & 44.64$\pm$13.8 \\
     & 2451884.74498 & 37.34$\pm$11.9 \\
     & 2451884.75713 & -12.76$\pm$11.9 \\
     & 2451898.68069 & 28.84$\pm$8.6 \\ 
     & 2451898.69285 & -11.86$\pm$9.1 \\ 
     & 2451898.70504 & -46.76$\pm$9.0 \\
\hline 
VLT$+$X-Shooter & 2455444.86040 &	-43.01$\pm$4.50 \\
     & 2455444.86933 &	-12.61$\pm$4.94 \\
     & 2455444.87819 &	16.90$\pm$4.72 \\
     & 2455444.88711	& 43.71$\pm$4.65 \\
\hline
HET$+$HRS & 2455536.74182 & -26.24$\pm$4.13 \\
  & 2455536.75732 & -59.02$\pm$6.34 \\
  & 2455540.72758 & -29.01$\pm$4.09 \\
  & 2455540.74309 & 19.97$\pm$6.01 \\
  & 2455577.62159 & 27.81$\pm$4.50 \\
  & 2455577.63707 & -9.44$\pm$4.58 \\
  & 2455584.61300 & 59.44$\pm$5.11 \\
  & 2455584.62852 & 51.29$\pm$8.52 \\
  & 2455600.57543 & -23.57$\pm$5.74 \\
\hline                             
\end{tabular}
\label{rvdata} 
\end{table}

Each SDSS spectrum is a combination of separate red and blue components. We split these into two datasets, the first covering H$\alpha$ and the second covering the higher Balmer series. We fitted multiple Gaussian components to these lines, using two Gaussians per line (see Marsh, Dhillon \& Duck 1995 for details of this process). Once profiles to the mean spectra had been fitted, then for the final fit we held all shape parameters fixed and simply allowed the radial velocity to be fitted, giving us our final radial velocities.

\begin{figure}
\centering
\includegraphics[width=5.9cm,angle=270]{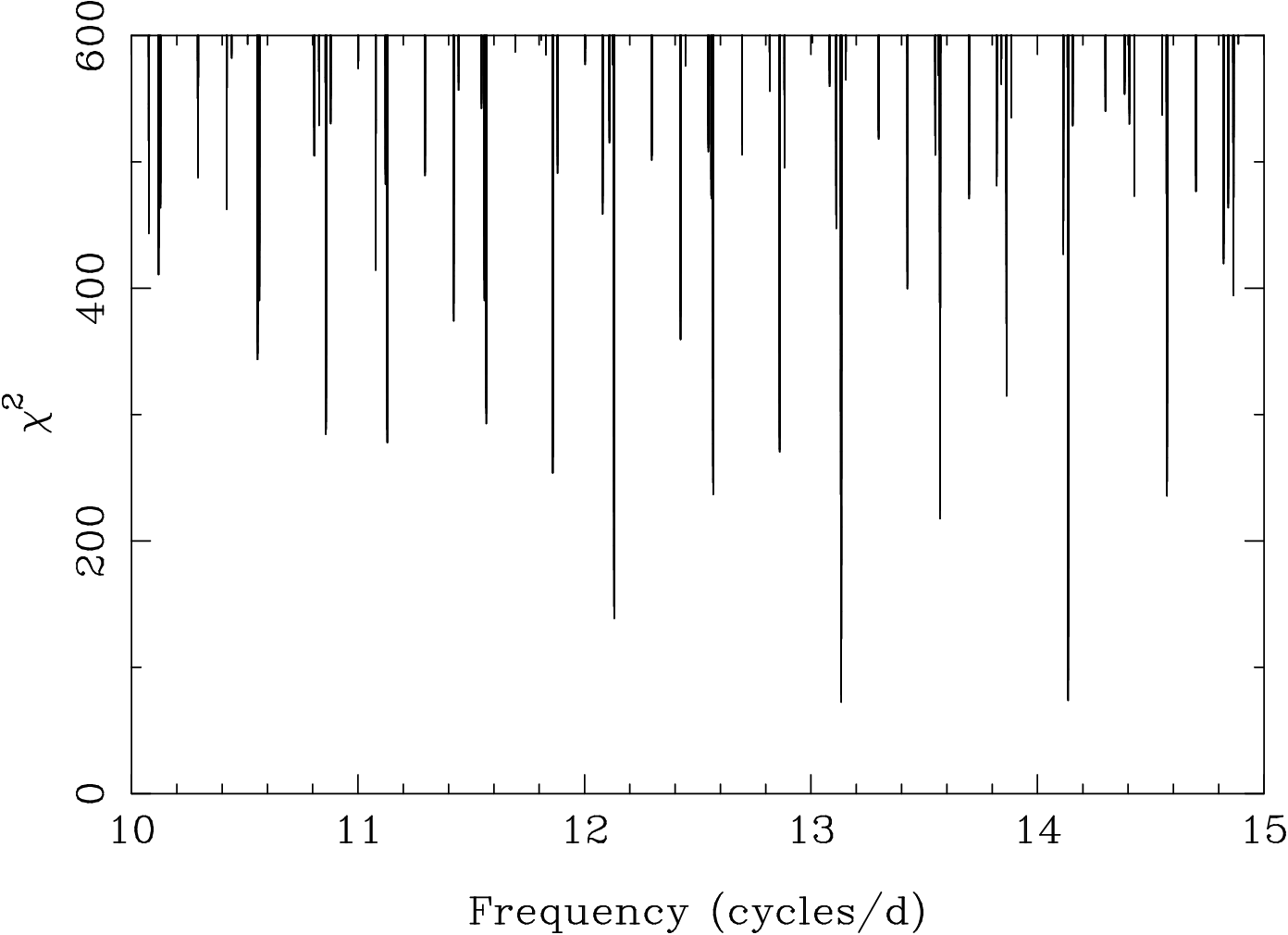}
\caption{Periodgram covering the region of interest for the WD's radial velocities as measured using the Balmer series absorption lines in the SDSS, HET and X-Shooter data. Two aliases are favoured at periods of 101.88$\pm0.02$ and 109.96$\pm$0.02 mins with a minimal difference in $\chi^{2}$.}
\label{rvfit_wd}
\includegraphics[width=5.9cm,angle=270]{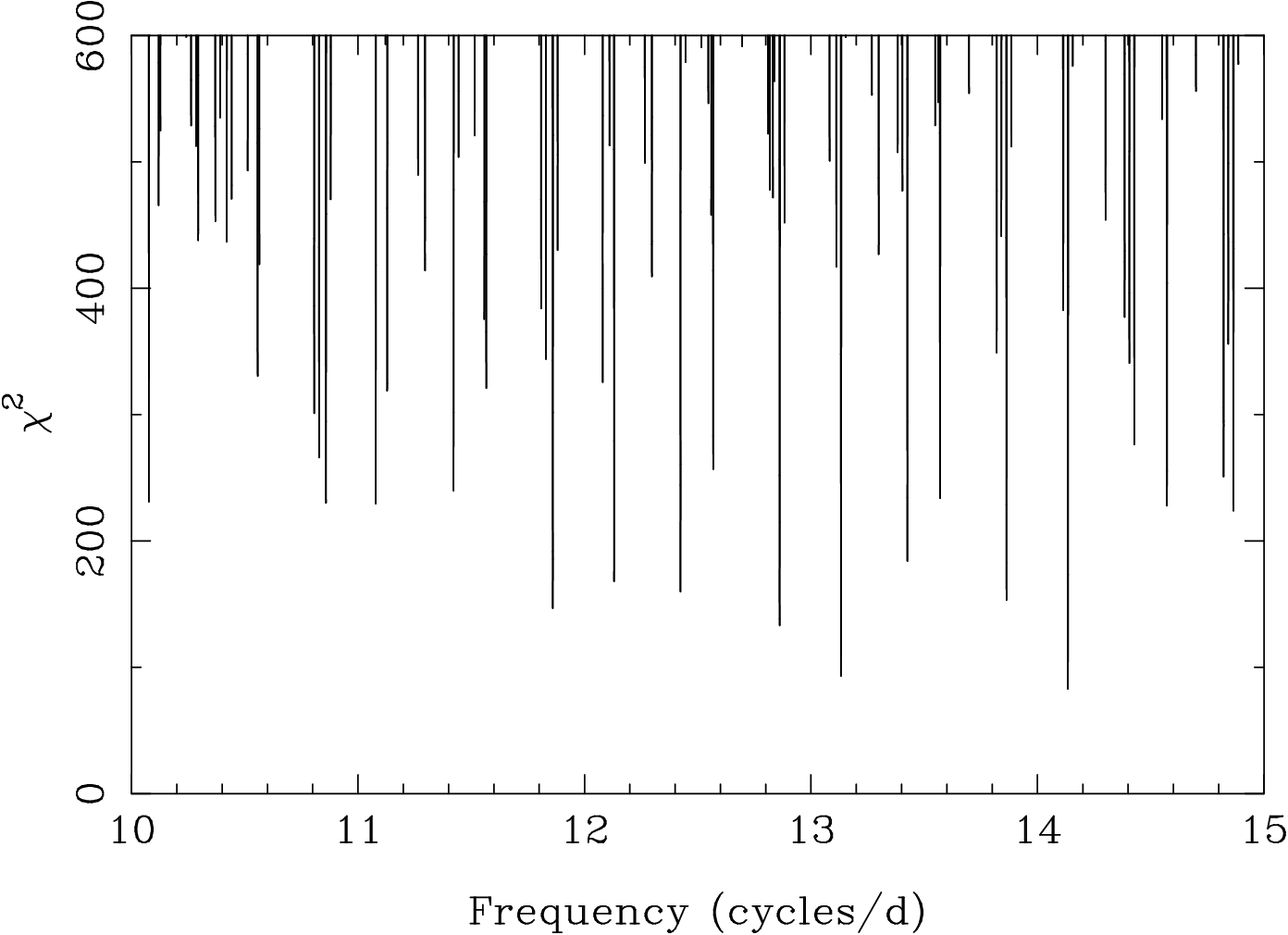}
\caption{Periodgram covering the region of interest for the BD's radial velocities as measured using the H$\alpha$ core emission in the HET and X-Shooter data. In this case the periodogram favours the higher frequency alias giving a period of 101.87$\pm$0.04 mins.}
\label{rvfit_bd}
\end{figure}

The HET spectra also consists of separate blue and red components, the blue end covering $H\beta$ and H$\gamma$, and the red covering H$\alpha$. Four of the five observations consisted of 2 separate exposures of 1320s each, with the final observation only producing one such exposure. As these exposures sampled a significant fraction of the orbital period of NLTT\,5306\,B, each observation was fitted separately, giving a total of 9 data points. Four more radial velocities were measured by extracting individually the spectra which were combined to make the final X-Shooter spectrum.



The absorption lines in the HET and X-Shooter data were fitted using a combination of 2 Gaussians using the program \textsc{FITSB2} written by Ralf Napiwotzki. The best fitting shape parameters were fixed and then fitted for velocity shifts. 

The final measured radial velocities for all 3 datasets are given in Table~\ref{rvdata}. The SDSS data were shifted so that the mean velocity matched that of the higher resolution and therefore superior HET and X-Shooter data. The low resolution of the SDSS data made measurement of the systemtic velocity with just the SDSS data unreliable, with a significant difference measured in the mean velocites ($\sim20$\,km/s) even between the blue and red arms. The mean velocity measured from HET and X-Shooter is consistent with the systemic velocity of -15$\pm$36 km/s previously measured in Adelman-McCarthy (2008).

The high resolution of HET HRS and X-Shooter spectra allowed for the detection of the line emission in the core of H$\alpha$. The fit to the WD's H$\alpha$ absorption line was used to subtract the contribution of the WD. The emission line was then fitted with a single Gaussian to measure the radial velocity. This was possible in all 4 X-Shooter exposures and 8 of the 9 HET exposures, with the 8th exposure being too noisy to accurately identify the emission.

\begin{figure}
\centering
\includegraphics[width=6.0cm,angle=270]{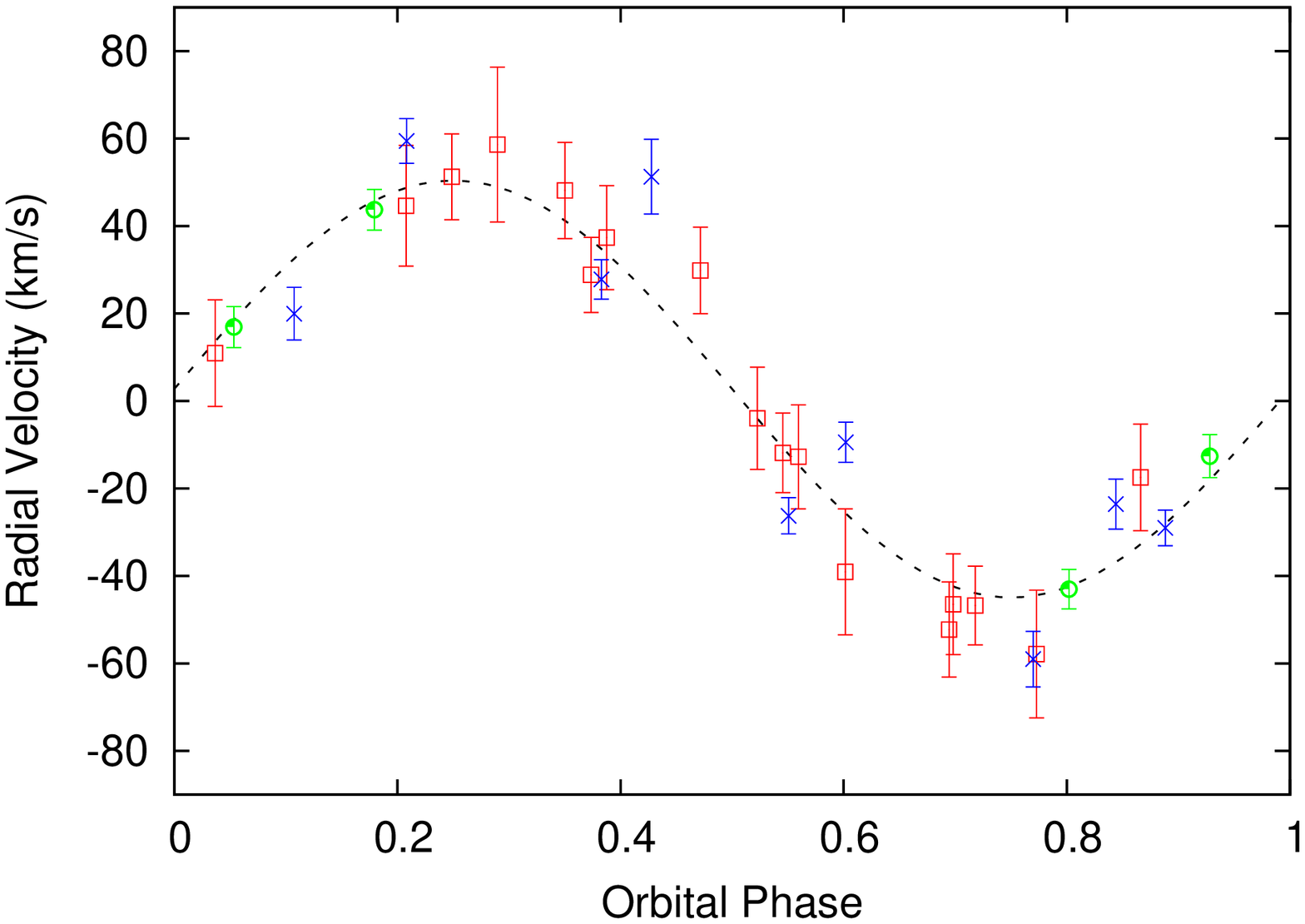}
\caption{Phase folded radial velocity curve for the favoured period of 101.87$\pm$0.04\,mins of the H absorption lines of NLTT\,5306\,A showing the SDSS (squares), HET (crosses) and X-Shooter (circles) data. The curve representing the best fitting orbital parameters is over-plotted (dashed).}
\label{wd_rv}
\includegraphics[width=6.0cm,angle=270]{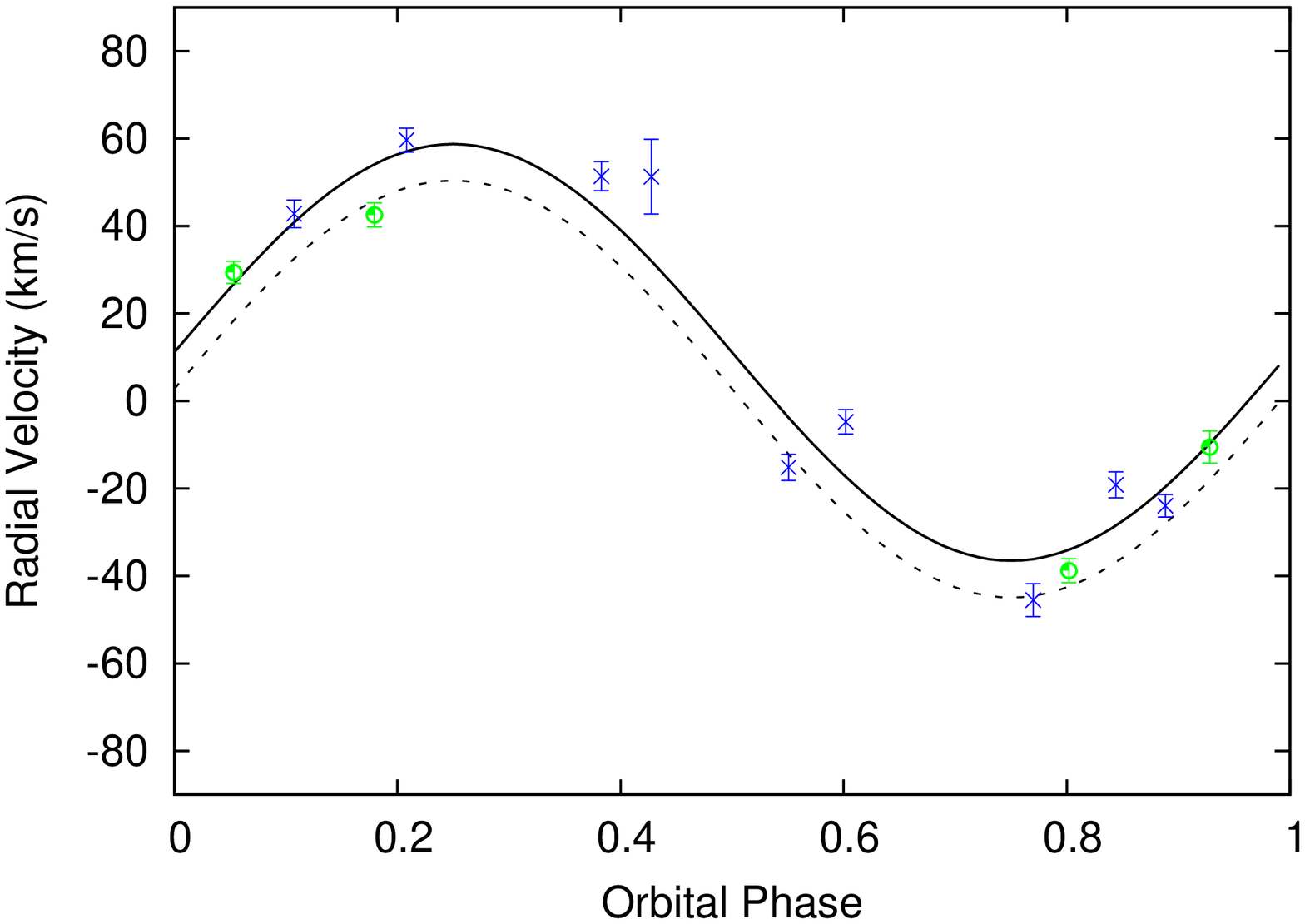}
\caption{Phase folded radial velocity curve for the favoured period of 101.87$\pm$0.04\,mins of the H$\alpha$ emission line showing the HET (crosses) and X-Shooter (circles) data. The curve representing the best fitting orbital parameters is over-plotted (solid) as well as the best fitting model for the absorption lines (dashed).}
\label{bd_rv}
\end{figure}

Periodicity was searched for in the combined radial velocity data (Table~\ref{rvdata}) of the WDs' absorption lines (SDSS, HET, and X-Shooter) and for the line emission (HET and X-Shooter) using the method discussed in Section~\ref{int}. For the WD, 30 unique data points were used as measured from the Balmer series absorption lines, and 12 as measured from the line emission in the core of H$\alpha$. 

Figure~\ref{rvfit_wd} shows the periodogram for the fit of the WD radial velocity measurements. The analysis of this data resulted in 2 favored aliases with periods of 109.96$\pm$0.02\,mins ($\chi^{2}=72.6$) and 101.88$\pm$0.02\,mins ($\chi^{2}=66.5$). The $\chi^{2}$ value slightly favors the shorter period alias which would be consistent with the period measured from the INT light curve. However, the longer period alias produces a period which is also within errors of this value. Therefore, it was not possible to determine the period accurately just utilising the available WD data.

Figure~\ref{rvfit_bd} shows the periodogram for the fit of the H$\alpha$ core emission radial velocity measurements. The 2 most favoured periods of 101.87$\pm$0.04\,mins ($\chi^{2}=78.5$) and 109.64$\pm$0.04\,mins ($\chi^{2}=93.0$) are indeed consistent with those obtained for both the light curve and the radial velocity fit of the WD. However, the $\chi^{2}$ value is much more in favour of the shorter period alias in this case. Therefore, we take the value of $P=$101.88$\pm$0.02\,mins from the larger (and therefore more accurate) absorption line dataset for the period of the system. 

An attempt was made to measure the radial velocity of the BD component using the four individual X-Shooter observations in the NIR arm. Each spectrum was extracted and flux calibrated, and the telluric correction applied. The model WD spectrum (Figure~\ref{xshooter_wd}) was then subtracted from each spectrum as in Section 3.3. We then cross correlated each spectrum with a dL5 template using the \textsc{IRAF} package \textsc{FXCOR}. This was attempted using various wavelength ranges in order to negate the water vapour bands. However the S/N in the individual exposures was such that the cross correlations with the template did not produce any sensible results. We also suspect that we were not entirely able to remove the WD contribution satisfactorily, and it is also possible that the BD spectrum undergoes short term changes in overall shape due to the irradiation from the primary.

\section{Discussion}

The stellar/substellar nature of an object is dependent on its mass \citep{kumar63}. The commonly used limit to distinguish between low mass main-sequence stars and BDs is 0.075\msun (75\mjup, \citealt{burrows97,chabbaraf00}), below which hydrogen fusion does not occur.

Figure~\ref{wd_rv} and \ref{bd_rv} show the phase folded radial velocity measurements for NLTT\,5306. The parameters of the spectroscopic orbit are summarised in Table~\ref{param_spec}. The period measured using the radial velocity data is consistent with the period measured using the variability of the INT $i'$-band light curve. Using the calculated value for the mass of the primary, $M_{\rm WD}=$0.44$\pm$0.04\,\msun (Table~\ref{param_wd}), the  minimum mass of the secondary is 56$\pm$3\mjup, consistent with the measured spectral type of dL4-dL7 and confirming NLTT\,5306\,B is a bona fide BD. Given this consistency, we suspect that this system has a relatively high inclination, even though no eclipse was detected. It also confirms NLTT\,5306 is the shortest period detached WD$+$BD binary. \cite{parsons12} detected a detached WD binary system with a period of only 94\,mins (CSS\,03170). However, in this case the secondary has a mass above the hydrogen burning limit and so is classed as a main sequence star. The most similar known system is WD\,0137$-$349 which has an orbital period of 116\,mins (Maxted et al. 2006) and a slightly lower mass secondary (53$\pm$6\mjup, Burleigh et al. 2006). 

One might assume that the H$\alpha$ emission seen in our spectroscopy is likely to arise from the irradiation of the BDs atmosphere by the WD primary, as is the case for WD\,0137$-$349. If this were true, then the radial velocity measurements would be in anti-phase with those measured from the H$\alpha$ absorption, and the amplitude would allow us to solve for the masses of both binary components. However, Figure~\ref{bd_rv} shows the emission is clearly in phase with the absorption, and with a similar measured amplitude of 48.9$\pm$1.8\,km/s, its origin must be associated with the WD. The most likely cause of such emission is accretion, either via Roche Lobe overflow or wind from the substellar companion.

Burleigh et al. (2006b) observed a similar situation in the magnetic WD$+$BD binary SDSS\,J\,121209.31$+$013627.7 ($P_{\rm orb}\approx90$\,mins). In this case the system is considered to be in a semi-detached state with a magnetic cataclysmic variable (polar) in a low state of accretion from a BD onto a magnetic WD. Although NLTT\,5306 shows no evidence of a detectable magnetic field and at $P_{\rm orb}\approx102$\,mins is probably not in semi-detached contact, it may be accreting from a weak wind from the brown dwarf. Therefore,  NLTT\,5306 is more akin to the wind accreting system LTT\,560 \citep{tappert07,tappert11}, albeit in this system the secondary is a much earlier spectral type of dM5. 

The H$\alpha$ emission seen in LTT\,560 consists of two anti-phased components, one originating in the secondary and the other from a chromosphere on the WD as a result of accretion via the companion's stellar wind. NLTT\,5306 only shows the emission component associated with the WD, and so we conclude we may be observing a similar situation where there is only chromospheric emission and no obvious activity from the secondary (which is to be expected given the estimated spectral type). The origin of this emission line component in post-common-envelope binaries (PCEB) is briefly discussed in \citet{tapgan11}, although given the rarity of systems where the H$\alpha$ emission line component is located on the WD, it is presently unclear under what conditions chomospheric emission occurs.

If NLTT\,5306 is accreting via a stellar wind onto a chromosphere then this would occur some distance above the WD. The systemic velocity of the observed emission is similar in value to that of the radial velocity measured using the WDs Balmer absorption lines. Some difference would be expected due to the gravitational redshift of the WD if the emission was chromospheric in origin. Adopting the measured atmospheric parameters from this work, this amounts to a redshift of $v_{\rm gr}=17.9\pm3.5$kms$^{-1}$. Figure~11 shows the systemic velocity of the emission feature is $\sim5-10$km/s greater than that measured from Balmer absorption. If anything we would expect it to be less (i.e redshifted compared to the WD) but given the low S/N of the emission detected in the HET data (from which the fit heavily relies), this value should not be trusted. In all likelihood the emission forming region is somewhere above the WDs photosphere. This could be further constrained with more accurate measurements of the radial velocity of the absorption and emission features, and should be considered as a future project for the VLT $+$ X-Shooter.


The effective temperature of the BD can be estimated to be $\sim1700$\,K from its measured spectral type of dL4-dL7, and comparison with observed L-dwarfs \citep{vrba04}. Using this effective temperature and the cooling age of the WD (Table~\ref{param_wd}) as a minimum value for the age of the system, we have estimated the radius of NLTT\,5306\,B to be $R_{\rm BD}=0.95\pm0.04$\,\Rjup\, by interpolating the Lyon group atmospheric models \citep{chabrier00,baraffe02}.

The systemic velocity of the WD allows us to discuss the kinematics of the system, in particular the $U$ velocity. This gives a good indication of whether the WD is a thin disc, thick disc or halo object, and thus allows for further constraints on the age of the primary. Using the equations of \citet{soderblom87} and the values given in Tables~\ref{param_wd} and \ref{param_spec} we calculate $U\approx70$\,km\,s$^{-1}$ for NLTT\,5306. This would seem to suggest that the WD is a member of the thick disk population (See Figure~4 of \citealt{pauli06}) and is likely much older than the minimum cooling age suggests ($>5$Gyr). At this age the Lyon group models give a mass of closer to 70\Mjup\ for the companion, still well within the accepted BD range.

The asymmetric heating and rotation of the BD produces a modulation of brightness known as the 'reflection effect' \citep{wilson90}. This has been detected at the order of $\sim1$\% in the INT $i'$-band light curve. Since the BD is tidally locked with the WD, this has allowed us to estimate the binary orbital period independently of the radial velocity measurements. A more accurate spectral typing of the companion would require further measurements of this effect at longer wavelengths (i.e. the near-infrared) where this variation would be more pronounced \citep{burleigh08}.

Variability has also been observed in the $i'$-band for WD\,0137$-349$, but of the order of $\sim2$\% (Burleigh, private communication). WD\,0137$-$349\,A is hotter than NLTT\,5306 ($\sim$16000\,K) so there are more UV photons, and more flux overall by approximately an order of magnitude. Therefore, we would not necessarily expect to observe such a strong effect on NLTT\,5306\,B, which may also depend on local conditions and chemistry in the BD atmosphere, but a variation of $\sim$1\% seems consistent with the effects of irradiation. 

The progenitor system of NLTT\,5306\,A\,\&\,B consisted of a main-sequence star and a BD with an orbital separation sufficiently small for the progenitor of NLTT\,5306\,A to fill its Roche lobe as it evolved off the main sequence. As a consequence of the ensuing unstable mass transfer the BD was engulfed in the envelope of the progenitor of NLTT\,5306\,A, leading to a rapid reduction in the orbital period and the ejection of the envelope. The low mass of NLTT\,5306\,A suggests that the core-growth was truncated by this common envelope evolution, and that this WD may contain a He-core \citep{webbink84-1, iben+tutukov86-1,rebassa-mansergasetal11-1}. Therefore, the evolution of NLTT\,5306\,A likely terminated on the red giant branch (RGB) rather the the asymptotic giant branch (AGB). 

Following the emergence from the common envelope, the binary continued to evolve towards shorter periods. Given the low mass of NLTT\,5306\,B gravitational wave radiation is likely to be the only relevant agent of orbital angular momentum loss. Adopting the stellar parameters for the WD and the BD determined above, and using the formalism outlined by \citet{schreiber+gaensicke03-1}, we calculate the orbital period at the end of the common envelope to have been $P_\mathrm{CE}\simeq120$\,min. The orbital period of NLTT\,5306 will continue to decrease for another $\simeq900$\,Myr, until the BD will eventually fill its Roche lobe and initiate stable mass transfer onto the WD. This transformation into a CV will occur at an orbital period of $\simeq68$\,min, near the orbital period minimum of CVs (G{\"a}nsicke et al. 2009).

The existence of WD0137-349\,B \citep{maxted06,burleigh06} and NLTT\,5306\,B demonstrate that BDs can survive common envelope evolution (see also \citealt{nordhaus10}), and their short orbital periods and low WD masses are in line with the statistics of the much larger sample of post-common envelope binaries containing low-mass M-dwarfs \citep{zorotovic11}. Binary population models predict both the existence of CVs born at very short periods with BD donors \citep{politano04-1}, and CVs containing low-mass He-core WDs \citep[e.g.][]{dekool92-1, politano96-1}. Yet, among the sample of known CVs, there is no compelling evidence for either systems that were born with a BD donor  (see the discussion in \citealt{littlefairetal07-1, uthasetal11-1, parsonsetal12-1,breedtetal12-2}) or CVs containing low-mass WDs \citep{zorotovicetal11-1, savouryetal11-1}. We conclude that WD0137-349 and NLTT\,5306 represent nearby bona-fide progenitors of CVs with low-mass WDs and brown-dwarf donors, and that the lack of such systems among the CV population reflects that the present-day population of pre-CVs is not fully representative of the progenitors of the present-day population of CVs.

\begin{table}
\caption{Properties of the white dwarf NLTT\,5306\,A} 
\label{param_wd} 
\centering   
\begin{tabular}{c c}
\hline    
Parameter  & Value \\                       
\hline
$T_{\rm eff}$ ($K$) & 7756$\pm$35 \\
log$g$ (c.g.s units) & 7.68$\pm$0.08 \\
Mass (\Msun) & 0.44$\pm$0.04 \\
Cooling Age (Myr) & 710$\pm$50 \\
Radius (\Rsun) & 0.0156$\pm$0.0016 \\
Distance (pc) & 71$\pm$4 \\
\hline                             
\end{tabular}
\end{table}

\begin{table}
\caption{Spectroscopic orbit of NLTT\,5306 where the WD radial velocity at a time $T$ is given by $\gamma_{1}+K_{1}\sin[2\pi f(T-T_{0})]$, and the emission line radial velocity $\gamma_{2}+K_{2}\sin[2\pi f(T-T_{0})]$, where $f=1/P$ is the frequency.} 
\label{param_spec} 
\centering   
\begin{tabular}{c c}
\hline    
Parameter  & Value \\                       
\hline
$P$\,(mins) & 101.88$\pm$0.02 \\
$T_{0}$ (HJD) & 2453740.1408$\pm$0.0005 \\
$K_{1}$ (km\,${\rm s}^{-1}$) & 48.1$\pm$1.3 \\
$K_{2}$ (km\,${\rm s}^{-1}$) & 48.9$\pm$1.8 \\
$\gamma_{1}$ (km\,${\rm s}^{-1}$) & 2.74$\pm$1.3 \\
$\gamma_{2}$ (km\,${\rm s}^{-1}$) & 11.10$\pm$1.0 \\
$a$ (\Rsun) & 0.566$\pm$0.005 \\
\hline                             
\end{tabular}
\end{table}

\section{Summary} 

We have spectroscopically confirmed the shortest period WD$+$BD binary known to date. Radial velocity variations and $i'$-band variability due to 'day' and 'night' side heating of the secondary give us a period of $101.88\pm0.02$\,mins, and a minimum mass for the companion of 56$\pm$3\mjup. This is consistent with the spectral type estimated from the spectroscopy of dL4-dL7. The results are summarised in Table~\ref{param_spec}. Emission near the core of H$\alpha$ indicates accretion either via a stellar wind or Roche Lobe overflow. NLTT\,5306\,B has survived a stage of common envelope evolution, much like its longer period counterpart WD\,0137$-$349. Both systems are likely to represent bona-fide progenitors of cataclysmic variables with a low mass white dwarf and a brown dwarf donor. 

\section{Acknowledgements}

PRS and MC are supported by RoPACS, a Marie Curie Initial Training Network funded by the European Commission's Seventh Framework Programme. RPS has received support from RoPACS during this research. MRB acknowledges receipt of an STFC Advanced Fellowship.
 
The Hobby-Eberly Telescope (HET) is a joint project of the University of Texas at Austin, the Pennsylvania State University, Stanford University, Ludwig-Maximilians-Universit\"at M\"unchen, and Georg-August-Universit\"at G\"ottingen. The HET is named in honor of its principal benefactors, William P. Hobby and Robert E. Eberly.

Based on observations made with the INT operated on the island of La Palma by the Isaac Newton Group in the Spanish Observatorio del Roque de los Muchachos of the Instituto de Astrofisica Canarias.

\bibliographystyle{mn}
\bibliography{wdplanets,browndwarfs,mbu,prs15,wd0132,btg}

\end{document}